\documentclass[aps,prl,twocolumn,showpacs,superscriptaddress,floatfix,nofootinbib]{revtex4}

\usepackage{graphicx}
\usepackage{amsmath}
\usepackage{epsfig}
\usepackage{helvet}
\usepackage{amssymb}

\newcommand{\be}{\begin{equation}}
\newcommand{\ee}{\end{equation}}
\newcommand{\bea}{\begin{eqnarray}}
\newcommand{\eea}{\end{eqnarray}}

\newcommand{\ket}[1]{\mbox{$| #1 \rangle$}}

\begin{document}

\title{Entanglement renormalization in two spatial dimensions}
\author{G. Evenbly}
\author{G. Vidal}
\affiliation{School of Mathematics and Physics, the University of Queensland, Brisbane 4072, Australia} 
\date{\today}

\begin{abstract}
We propose and test a scheme for entanglement renormalization capable of addressing large two-dimensional quantum lattice systems. In a translationally invariant system, the cost of simulations grows only as the logarithm of the lattice size; at a quantum critical point, the simulation cost becomes independent of the lattice size and infinite systems can be analysed. We demonstrate the performance of the scheme by investigating the low energy properties of the 2D quantum Ising model on a square lattice of linear size $L=\{ 6,9,18,54, \infty\}$ with periodic boundary conditions. We compute the ground state and evaluate local observables and two-point correlators. We also produce accurate estimates of the critical magnetic field and critical exponent $\beta$. A calculation of the energy gap shows that it scales as $1/L$ at the critical point.  
\end{abstract}

\pacs{05.30.-d, 02.70.-c, 03.67.Mn, 05.50.+q}

\maketitle

Entanglement renormalization \cite{ER} has been recently proposed as a real-space renormalization group (RG) method \cite{Wilson} to study extended quantum systems on a lattice. A highlight of the approach is the removal, before the coarse-graining step, of short-range entanglement by means of unitary transformations called \emph{disentanglers}. This prevents the accumulation of short-range entanglement over successive RG transformations. Such accumulation is the reason why the density matrix renormalization group (DMRG) \cite{DMRG}-- an extremely powerful technique for lattices in one spatial dimension -- breaks down in two dimensions, where it can only address small systems. 

The use of disentanglers leads to a real-space RG transformation that can in principle be iterated indefinitely, enabling the study of very large systems in a quasi-exact way. This RG transformation also leads to the so-called \emph{multi-scale entanglement renormalization ansatz} (MERA) \cite{MERA} to describe the ground state of the system -- or, more generally, a low energy sector of its Hilbert space. In a translation invariant lattice made of $N$ sites, the cost of simulations grows only as $\log N$ \cite{NewAlgorithm}. In the presence of scale invariance, this additional symmetry is naturally incorporated into the MERA and a very concise description, independent of the size of the lattice, is obtained in the infrared limit of a topological phase \cite{topo} or at a quantum critical point \cite{ER, MERA, FreeFermions, FreeBosons, Transfer, CFT, Transfer2}.
 
While the basic principles of entanglement renormalization are the same in any number of spatial dimensions, most available calculations refer to 1D models. Numerical work with 2D lattices incurs a much larger computational cost and has so far been limited to exploratory studies of free fermions \cite{FreeFermions} and free bosons \cite{FreeBosons} and of the Ising model in a square lattice of small linear size $L\leq 8$ \cite{Finite2D}. It must be emphasized, however, that the approach of Refs. \cite{FreeFermions, FreeBosons} relies on the gaussian character of free particles and can not be generalised to the interacting case, whereas the results of Ref. \cite{Finite2D} were obtained by exploiting a significant reduction in computational cost that occurs only for small 2D lattices. 

In this paper we present an implementation of the MERA that allows us to consider, with modest computational resources, 2D systems of arbitrary size, including infinite systems. In this way we demonstrate the scalability of entanglement renormalization in two spatial dimensions and decisively contribute to establishing the MERA as a competitive approach to systematically address 2D lattice models. The key of the present scheme is a carefully planned organization of the tensors in the MERA, leading to simulation costs that grow as $O(\chi^{16})$, where $\chi$ is the dimension of the vector space of an effective site. This is drastically smaller than the cost $O(\chi^{28})$ of the best previous scheme \cite{FreeFermions,FreeBosons,Finite2D}. We also demonstrate the performance of the scheme by analysing the 2D quantum Ising model, for which we obtain accurate estimates of the ground state energy and magnetizations, as well as two-point correlators (shown to scale polynomially at criticality), the energy gap, and the critical magnetic field and beta exponent. Finally, we discuss how the use of disentanglers affects the simulation costs, by comparing the MERA with a \emph{tree tensor network} (TTN) \cite{TTN}.

\begin{figure}[!tb]
\begin{center}
\includegraphics[width=8cm]{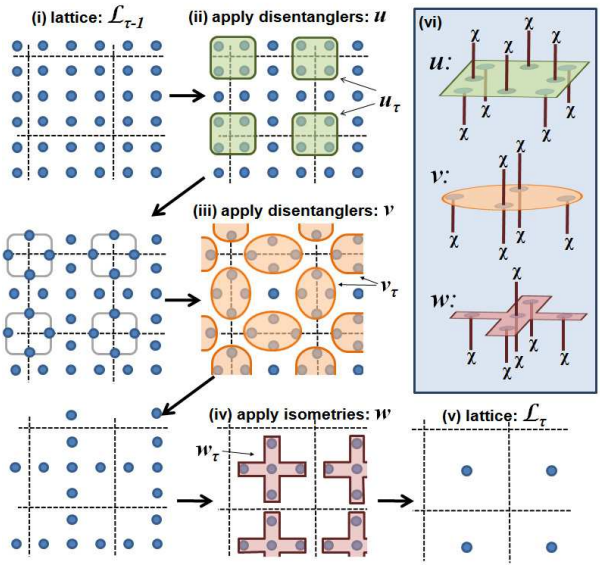}
\caption{(Color online) Entanglement renormalization scheme for a square lattice. A block of $3\times 3$ sites of lattice $\mathcal{L}_{\tau-1}$ (i) is mapped onto one site of $\mathcal{L}_{\tau}$ (v). The RG transformation involves (ii) applying disentanglers $u$ between the corners of adjacent blocks followed by (iii) disentanglers $v$ which act across the sides of adjacent blocks and (iv) isometries $w$ which act within a block. Tensors $u,v$ and $w$ have a varying number of incoming and outgoing indices (vi) according to Eq. \ref{eq:tensors}.} \label{fig:Scheme} 
\end{center}
\end{figure}
 
\textbf{2D MERA.---} 
Let us consider a square lattice $\mathcal{L}_0$ made of $N=L\times L$ sites, each one described by a Hilbert space $\mathbb{V}$ of finite dimension $d$. The proposed 2D MERA is characterized by the coarse-graining transformation of Fig. (\ref{fig:Scheme}), where blocks of $3 \times 3$ sites of lattice $\mathcal{L}_{0}$ are mapped onto single sites of a coarser lattice $\mathcal{L}_1$. This is achieved in three steps: first disentanglers $u$ are applied on the four sites located at the corners of four adjacent blocks; then disentanglers $v$ are applied at the boundary between two adjacent blocks, transforming four sites into two; finally, isometries $w$ are used to map a block into a single effective site. In this way, tensors $u,v$ and $w$ \cite{isometric},  
\begin{equation}
	u^{\dagger}\!: \mathbb{V}^{\otimes 4} \rightarrow \mathbb{V}^{\otimes 4},~~~
	v^{\dagger}\!: \mathbb{V}^{\otimes 4} \rightarrow \mathbb{V}^{\otimes 2},~~~
  w^{\dagger}\!: \mathbb{V}^{\otimes 5} \rightarrow \mathbb{V}, \label{eq:tensors}
\end{equation}
transform the state $\ket{\Psi_0} \in \mathbb{V}^{\otimes N}$ of the lattice $\mathcal{L}_0$ in which we are interested (typically the ground state of a local Hamiltonian $H_0$) into a state $\ket{\Psi_1}\in \mathbb{V}^{\otimes N/9}$ of the effective lattice $\mathcal{L}_1$ through the sequence 
\begin{equation}
	\ket{\Psi_0} \stackrel{u}{\rightarrow} \ket{\Psi_0'} \stackrel{v}{\rightarrow} \ket{\Psi_0''} \stackrel{w}{\rightarrow} \ket{\Psi_1}.
\end{equation}
To understand the role of these tensors, it is useful to think of the state $\ket{\Psi_0}$ as possessing three different kinds of entanglement: short-range entanglement residing at the corners of four adjacent blocks, short-range entanglement residing near the boundary shared by two blocks, and long-range entanglement. Then the disentanglers $u$ and $v$ are used to reduce the amount of short-range entanglement residing near the corners and boundaries of the blocks. In other words, in states $\ket{\Psi_0'}$ and $\ket{\Psi_0''}$ increasing amounts of short-range entanglement from $\ket{\Psi_0}$ have been removed. This fact facilitates significantly the job of the isometry $w$, namely to compress into an effective site of $\mathcal{L}_1$ those degrees of freedom in a block that still remain entangled (now mostly through long-range entanglement) with degrees of freedom outside the block. Thus, the resulting state $\ket{\Psi_1}$ still contains the long-range entanglement of $\ket{\Psi_0}$, but most of its short-range entanglement is gone. We complete the above construction by noticing that a $d$-dimensional space $\mathbb{V}$ is often too small to accommodate all the relevant degrees of freedom left on a block. Accordingly, we shall describe the effective sites of $\mathcal{L}_1$ with a space of larger dimension $\chi$. This dimension $\chi$ determines both the accuracy and cost of the simulations. 

The transformation of Fig. \ref{fig:Scheme} can now be applied to lattice $\mathcal{L}_1$, producing a coarser lattice $\mathcal{L}_2$. More generally, if $\mathcal{L}_0$ is finite, $O(\log N)$ iterations will produce a sequence of lattices $\{\mathcal{L}_0, \mathcal{L}_1, \mathcal{L}_2,\cdots , \mathcal{L}_{\mbox{\tiny top}} \}$ where the top lattice $\mathcal{L}_{\mbox{\tiny top}}$ contains only a small number of sites and can be addressed with exact numerical techniques. Thus, given a Hamiltonian $H_0$ on $\mathcal{L}_{0}$, we can use the above RG transformation to obtain a sequence of Hamiltonians $\{H_0, H_1, H_2, \cdots, H_{\mbox{\tiny top}}\}$, then diagonalize $H_{\mbox{\tiny top}}$ to find its ground state $\ket{\Psi_{\mbox{\tiny top}}}$, and finally recover the ground state $\ket{\Psi_0}$ of $H_0$ by reversing all the RG transformations:
\begin{equation}
	\ket{\Psi_{\mbox{\tiny top}}} \rightarrow \cdots \rightarrow \ket{\Psi_2} 
\rightarrow \ket{\Psi_1} \rightarrow \ket{\Psi_0}.
\label{eq:sequence}
\end{equation}
This is precisely how the MERA is defined. Specifically, the MERA for $\ket{\Psi_0}$ is a tensor network containing (i) a top tensor, that describes $\ket{\Psi}_{\mbox{\tiny top}}$, and (ii) $O(\log N)$ layers of tensors (disentanglers and isometries), where each layer is used to invert one step of the coarse-graining transformation of Fig. \ref{fig:Scheme} according to the sequence (\ref{eq:sequence}).

\begin{figure}[!tb]
\begin{center}
\includegraphics[width=8cm]{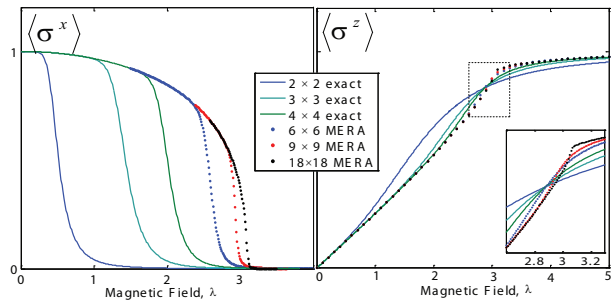}
\caption{(Color online) Spontaneous and transverse magnetizations $\left\langle {\sigma _x } \right\rangle$ and $\left\langle {\sigma _z } \right\rangle$ as a function of the applied magnetic field $\lambda$ and for different lattice sizes $L$. Results for small systems correspond to exact diagonalization whilst results for larger systems were obtained with a $\chi=6$ MERA. As $L$ increases, the magnetizations are seen to converge toward their thermodynamic limit values. Results for $L=54$ could not be visually distinguished from results for $L=18$ and have been omitted in the plot. As it is characteristic of a second order phase transition, for large $L$ both magnetizations develop a discontinuity in their derivative, with $\left\langle {\sigma _x } \right\rangle$ (the order parameter) suddenly dropping to zero at the quantum critical point (see Fig. \ref{fig:MagRefine}).}\label{fig:MagScale} 
\end{center}
\end{figure}

The technical details on how to numerically optimize the disentanglers and isometries of the MERA to approximate the ground state $\ket{\Psi_0}$ of $H_0$ are analogous to those discussed in Ref. \cite{NewAlgorithm} for a 1D lattice and will not be repeated here. Instead, we focus on the key aspect that makes the present 2D scheme much more efficient than that of Refs. \cite{FreeFermions,FreeBosons,Finite2D}. For this purpose, we consider an operator $O_0$ whose support is contained within a block of $2\times 2$ sites of lattice $\mathcal{L}_0$. Direct inspection shows that, no matter where this block is placed with respect to the disentanglers and isometries of Fig. \ref{fig:Scheme}, the support of the resulting coarse-grained operator $O_1$ is also contained within a block of $2\times 2$ sites of $\mathcal{L}_1$, and the same holds for any subsequent coarse-graining. This is in sharp contrast with the 2D scheme of Refs. \cite{FreeFermions,FreeBosons,Finite2D}, where the minimal stable support of local observables (or 'width' of past \emph{causal cones}) corresponded to blocks of $3\times 3$ sites. In the present case, much smaller objects (operators acting on $4$ sites instead of $9$ sites) are manipulated during the calculations, resulting in the announced dramatic drop in simulation costs.

\begin{figure}[!tp]
\begin{center}
\includegraphics[width=8cm]{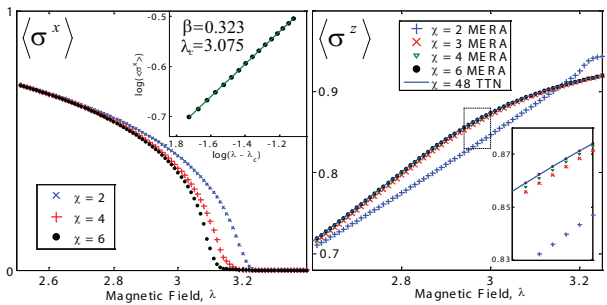}
\caption{(Color online) Magnetizations $\left\langle {\sigma _x } \right\rangle$ and $\left\langle {\sigma _z } \right\rangle$ as a function of the applied magnetic field $\lambda$ for different values of the refinement parameter $\chi$. \emph{Left:} 
Spontaneous magnetization $\left\langle {\sigma _x } \right\rangle$ for $L=54$. Data fits of the form $\left\langle {\sigma _x } \right\rangle  \sim \left( {\lambda  - \lambda _c } \right)^{\beta_c }$ near the critical point give a critical magnetic field $\lambda _c  = \left\{ {3.13,3.09,3.075} \right\}$ and critical exponent $\beta _c  = \{ 0.320,0.321,0.323\}$ for $\chi  = \left\{ {2,4,6} \right\}$. Current Monte Carlo estimates are $\lambda _c  = 3.044$ and $\beta _c  = 0.326$ \cite{MC}. Thus accuracy increases with $\chi$. \emph{Right:} Transverse magnetization $\left\langle {\sigma _z } \right\rangle$ for $L=6$. TTN results for large $\chi$ are taken as the exact solution (see Fig. \ref{fig:2DEnergyErrorNew}). Whilst a $\chi=2$ MERA produces significantly different values, results for $\chi=3$ are already very similar and those for $\chi=6$ MERA agree with the TTN solution on at least 3 significant digits.} \label{fig:MagRefine}
\end{center}
\end{figure}


\textbf{Benchmark calculations.---} We have tested the proposed scheme by investigating low energy properties of the quantum Ising model with transverse magnetic field,
\begin{equation}
H_{\mbox{\tiny Ising}} = \sum\limits_{\left\langle {r,r'} \right\rangle } {\sigma _x^{[r]} \sigma _x^{[r']}  + \lambda \sum\limits_r {\sigma _z^{[r]} } },
\end{equation}
on a square lattice with periodic boundary conditions (local dimension $d=2$). 
First of all, we consider a sequence of lattices with increasing linear size $L=\{6,9,18,54\}$. For each of them, a MERA approximation to the ground state of $H_{\mbox{\tiny Ising}}$ for different values $\lambda \in [0, 5]$ of the transverse magnetic field is obtained using $\chi=6$. Computing the ground state for $L=54$ and critical transverse magnetic field takes $\sim$ 4 days on a 3GHz dual-core desktop PC with 8Gb RAM when starting from a randomly initialized MERA \cite{chireduction}. Fig. \ref{fig:MagScale} displays the expected value of the parallel and transverse magnetizations, both of which show characteristic signs of a second order phase transition as $L$ increases. We emphasize that since the simulation costs grow only as the logarithm of $L$, it is straightforward to increase the system size until e.g. finite size effects become negligible on local observables.

Fig. \ref{fig:MagRefine} shows how the parallel and transverse magnetizations change with increasing $\chi$, for $L=54$. Since the cost of the simulations grows as $O(\chi^{16})$, only small values of $\chi$ can be considered in practice. However, with $\chi=6$ one already obtains estimates for the location of the critical point and the critical exponent $\beta$ that already fall within $1\%$ of the best Monte Carlo results \cite{MC}.

\begin{figure}[!tbp]
\begin{center}
\includegraphics[width=8cm]{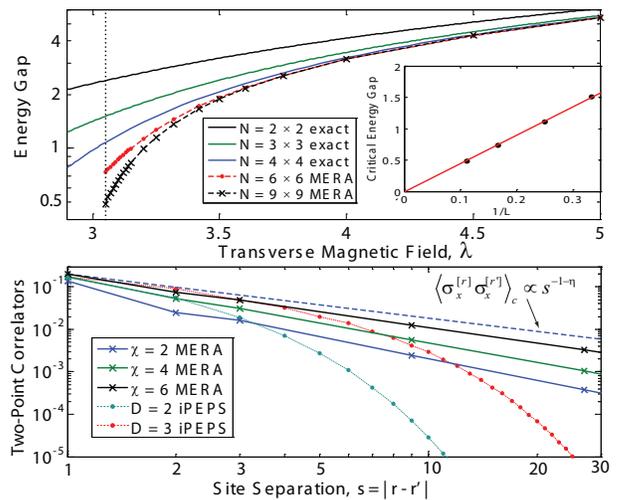}
\caption{(Color online) \emph{Top:} The energy gap as a function of the transverse magnetic field $\lambda$, computed by exact diagonalization for small system sizes $L=\left\{2,3,4\right\}$ and with a $\chi=6$ MERA for $L=\left\{6,9\right\}$. The gap scales as $1/L$ at the critical magnetic field. \emph{Bottom:} Two-point correlators $\langle {\sigma_x^{[r]} \sigma_x^{[r']} } \rangle_{c}$ at criticality and for different values of $\chi$. The scale invariant MERA produces correlators that decay polynomially with the distance $s \equiv |r-r'|$. As $\chi$ increases their asymptotic scaling approaches $1/s^{ 1 + \eta }$ with $\eta = 0.03 \pm 0.01$ \cite{critphenom}. Correlators have been computed at distances $s=3^k$ for $k=0,1,2,\ldots$, where they can be evaluated with cost $O(\chi^{16})$. For comparison, we have included correlators obtained with a $D=2$ and $D=3$ iPEPS \cite{iPEPS}. The latter are very accurate for $s=1,2$ but decay exponentially after a few sites. } \label{fig:GapCorr}
\end{center}
\end{figure}

By using the MERA to represent a two-dimensional subspace and minimizing the expectation value of $H_{\mbox{\tiny Ising}}$, we obtain the system's energy gap $\Delta E$. Fig. \ref{fig:GapCorr} shows $\Delta E$ as a function of the transverse magnetic field and system size. Notice that at the critical point the gap closes with the system size as $1/L$ (dynamic exponent $z=1$). Two-point correlators can also be extracted. Fig. \ref{fig:GapCorr} shows the correlator $\langle \sigma_{x}^{[r]} \sigma_{x}^{[r']} \rangle_{c} \equiv \langle \sigma_{x}^{[r]} \sigma_{x}^{[r']} \rangle - \langle \sigma_{x}^{[r]} \rangle\langle \sigma_{x}^{[r']}\rangle$ along a row or column of the lattice, obtained using the scale invariant algorithm \cite{CFT}, which directly addresses an infinite lattice at the critical point. 

\begin{figure}[!t]
\begin{center}
\includegraphics[width=8cm]{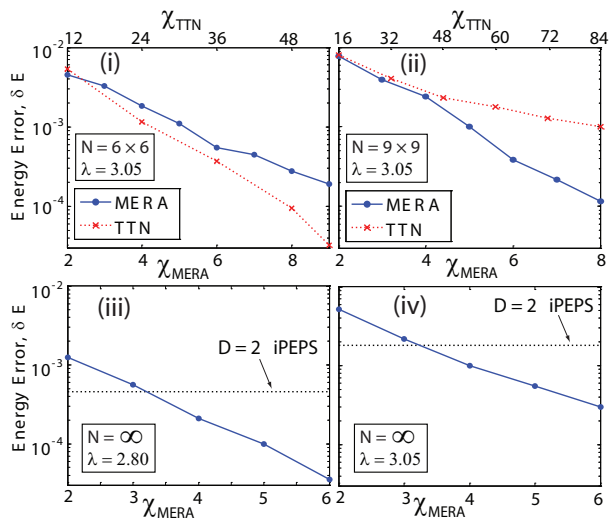}
\caption{(Color online) Energy error as a function of the refinement parameter $\chi$ for finite systems of different sizes and for infinite systems. In absence of an exact solution for ground state energies, the errors are defined relative to the results obtained with (i) a $\chi=60$ TTN, (ii) a $\chi=9$ MERA, (iii,iv) a $D=3$ iPEPS \cite{iPEPS}. For finite systems (i,ii), the MERA is compared against the TTN. The double $x$-axes for $\chi_{\mbox{\tiny MERA}}$ and $\chi_{\mbox{\tiny TTN}}$ have been adjusted so that they roughly correspond to the same computational cost. For $L=6$ the TTN is more efficient whilst for $L=9$ the MERA already gives significantly better results. Comparison between MERA and iPEPS results for (iii) an infinite system off criticality and (iv) an infinite system at criticality shows very similar accuracy between $\chi=3$ MERA and $D=2$ iPEPS, whereas $D=3$ iPEPS gives a lower (better) energy than $\chi=6$ MERA.} \label{fig:2DEnergyErrorNew} 
\end{center}
\end{figure}

\textbf{Role of disentanglers.---} In order to highlight the importance of disentanglers, we have also performed simulations with a tree tensor network (TTN). This corresponds to a more orthodox real-space RG approach where the block of $3\times 3$ sites in Fig. \ref{fig:Scheme} is directly mapped into an effective site without the use of disentanglers. Recall that a 2D ground state typically displays a boundary law, $S_l \approx l$, for the entanglement entropy $S_l$ of a block of $l\times l$ sites. To reproduce this boundary law with a TTN, one needs to increase the dimension $\chi$ at each step of the coarse-graining. Specifically, $\chi_{\mbox{\tiny{TTN}}}$ must grow doubly exponentially with the linear size $L$ of the lattice. On the other hand, the cost of manipulating a 2D TTN grows only as a small power of $\chi_{\mbox{\tiny{TTN}}}$. As a result, much larger values of $\chi$ can be used with a TTN, leading to a very competitive approach for small lattice sizes \cite{TTN}. Fig \ref{fig:2DEnergyErrorNew} (i and ii) compares the performance of the MERA and the TTN in lattices of size $6 \times 6$ and $9 \times 9$. It shows that a TTN is more efficient than the MERA in computing the ground state of the $6\times 6$ lattice; however, this trend is already reversed in the  $9\times 9$ lattice, where the cumulative benefit of using disentanglers clearly outweighs the large cost they incur. Disentanglers, by acting on the boundary of a block, readily reproduce the entropic boundary law (for any value of $\chi$) and allow us to consider arbitrarily large systems. Fig. \ref{fig:2DEnergyErrorNew} (iii and iv) shows results for an infinite lattice near and at criticality.

To summarize, we have proposed an entanglement renormalization scheme for the square lattice and demonstrated its scalability by addressing the quantum Ising model on systems of linear size $L=\{6,9,18,54\}$, with cost $O(\chi^{16} \log L)$, and on an infinite system at criticality, with cost $O(\chi^{16})$. The key of the present approach is the use of two types of disentanglers that remove short-range entanglement residing near the corners and near the boundaries of the blocks while leading to narrow causal cones of $2\times 2$ sites. Similar schemes can be built e.g. for triangular, hexagonal and Kagome lattices \cite{kagMERA}.

The authors thank Roman Orus, Luca Tagliacozzo and Philippe Corboz for comments. Support from the Australian Research Council (APA, FF0668731, DP0878830) is acknowledged.


\begin{thebibliography}{99}
\bibitem{isometric} The tensors of the MERA are called disentanglers $u$ or isometries $w$ depending on whether they are in charge of eliminating short-range entanglement or of mapping a block of sites into a single site \cite{ER,MERA,NewAlgorithm}. This distinction is somewhat arbitrary: one can consider tensors that fulfill the two roles simultaneously, such as tensor $v$ in Fig. \ref{fig:Scheme}, that we still call disentangler. All these tensors must be isometric, that is $u^{\dagger}u=I$, $v^{\dagger}v=I$, $w^{\dagger}w=I$. The hermitian conjugation ($\dagger$) in Eq. (\ref{eq:tensors}) appears for consistency with previous references.
\bibitem{chireduction} Calculations for $\chi=6$ are achieved by using a disentangler $u$ with $\chi=4$ on selected indices. The computation time is reduced to a few hours per point by re-using a MERA previously converged (for a similar magnetic field) as the starting point of a simulation.
\end{thebibliography}
\end{document}